\documentclass[12pt]{article}
\def\beq{\begin{equation}}
\def\eeq{\end{equation}}
\def\bea{\begin{eqnarray}}
\def\eea{\end{eqnarray}}
\begin{document}
\rightline{ NPI MSU  2001-21/661}
\rightline{May 2001}
\vspace{1cm}
\begin{center}
{\Large \bf Effective Lagrangians for linearized gravity \\
 in Randall-Sundrum model } \\

\vspace{4mm}

Edward E. Boos, Yuri A. Kubyshin,  Mikhail N. Smolyakov$^a$,
  Igor P.~Volobuev\\ Institute of Nuclear
Physics, Moscow State University \\ 119899 Moscow \\
 Russia \\
$^a$ Physical Department,  Moscow State University \\ 119899 Moscow \\
 Russia \\
\end{center}

\begin{abstract}
We construct the second variation Lagrangian for Randall-Sundrum model
with two branes, study its gauge invariance, find the corresponding
equations of motion and decouple them. We also derive an effective Lagrangian
for this
model in the unitary gauge, which describes a massless graviton, massive
gravitons and a massless scalar radion. It is shown that the mass
spectra of these fields  and their  couplings to the matter are different
on different branes.
\end{abstract}

\section{Introduction}
 The Kaluza-Klein (KK) hypothesis has been discussed in
theoretical physics for three quarters of a century. In accordance with
this hypothesis, space-time may have extra dimensions, which are
unobservable for certain reasons. The explanation of this unobservability,
which was put forward in the original papers by Kaluza and Klein, implies
that the extra dimensions are compactified and have a very small size of
the order of the Planck length $l_{Pl} = 1/M_{Pl}$.

In 1983 Rubakov and Shaposhnikov put forward a new scenario for Kaluza-Klein
theories, which was based on the idea of localization of fields on a domain
wall \cite{RSH}. They have also proposed an ansatz for multidimensional
metric,  which is compatible with this hypothesis  \cite{RSH1}.

In the last years there appeared indications that { scenarios of this type
can arise} in the theory of strings \cite{string} (see Ref. \cite{Ant01}
for a review and references). In this approach { our three spatial
dimensions are supposed to be realized as a three-dimensional hypersurface
embedded into a multidimensional space-time. Such hypersurfaces are called
3-branes}, or just branes.
 The main goal of such scenarios was to find a solution to the hierarchy
problem. It was solved either due to the sufficiently large
characteristic size of extra dimensions \cite{ADD}, or due to
exponential warp factor appearing in the metric \cite{RS1,RS2}.
In both approaches  gravity in multidimensional space-time becomes "strong"
not at the energies of the order of  $10^{19}\,$GeV, but at much lower
energies, maybe of the order of  $1 \div 10\,$TeV.
An attractive feature of these models is that they predict new effects
which can be observed at the coming collider experiments.

In paper  \cite{RS1} an exact solution for a system of two branes
interacting with gravity in five-dimensional space-time was found, which
allows one to study the effect of  the  gravitational field induced by the
branes. This model is called Randall-Sundrum model, and it was widely
discussed in the literature (see  Refs. \cite{Ant01,Rub01}
for reviews and references).  However,
there are some issues which are still to be clarified.
In the present paper we present a detailed derivation of the
Lagrangian of quadratic fluctuations around the Randall-Sundrum solution
and discuss its properties. In particular, we carry out a
diagonalization of this Lagrangian and of the corresponding equations
of motion.

The paper is organized as follows. In Sect.~2 we derive an expression for
the Lagrangian of quadratic fluctuations around an arbitrary background
convenient for our purposes. In Sect.~3 this result is employed for the
calculation of the Lagrangian of quadratic fluctuations in the
Randall-Sundrum model. We also  discuss here the gauge transformations in
the model and identification of physical degrees of freedom. In Sect.~4 we
calculate the couplings of the fields, arising from the five-dimensional
metric, to the matter. We emphasize that for the proper interpretation of
the interactions Galilean coordinates on the branes should be used. Sect.~5
contains some discussion of the effective theories on the branes.

\section{Second variation Lagrangian}
First, let us consider the standard gravitational action with cosmological
constant in $4+d$-dimensional space-time $E$ with coordinates $\{x^M\},
M=0,1, \cdots d+3$:
\begin{equation}\label{actiong}
S_g = \frac{1}{16 \pi \hat G} \int_E \left(R-\Lambda\right)
\sqrt{-g}\, d^{4+d}x,
\end{equation}
where $\hat{G}$ is the multidimensional gravitational constant.
The signature of the metric $g_{MN}$ is chosen to be $(-1,1, \cdots 1)$.
Let $\gamma_{MN}$ be a fixed background metric. We denote
$\hat \kappa = \sqrt{16 \pi \hat G}$ and parameterize  the metric  $g_{MN}$ as
\begin{equation}\label{metricpar}
  g_{MN} = \gamma_{MN} + \hat \kappa h_{MN}.
\end{equation}

If we substitute this formula into (\ref{actiong}) and retain only the terms
of the zeroth order in $\hat \kappa$, we
get the following Lagrangian, which is  usually called the second variation
Lagrangian:
\begin{eqnarray}\label{Lagrangian0}
L^{(2)}_g/\sqrt{-\gamma} & =& -\frac{1}{4}\left(\nabla_R h_{MN} \nabla^R h^{MN} -
\nabla_R h \nabla^R h + 2\nabla_M h^{MN}\nabla_N h -\right.
 \\ \nonumber
&-& \left. 2 \nabla^R h^{MN} \nabla_M h_{RN}\right) +
\frac{1}{4}\left(R-\Lambda \right)\left(h_{MN} h^{MN} -\frac{1}{2}
hh\right) +
\\ \nonumber
&+&  G^{MN} h_{MR} h_N^R - \frac{1}{2}G^{MN} h_{MN} h.
\end{eqnarray}
Here $\gamma = \mbox{det} \gamma_{MN} $, $h = h_M^M$,
 $ \nabla_M$ denotes the covariant derivative with respect
to metric $ \gamma_{MN}$, and
 $G_{MN}=  R_{MN} - \frac{1}{2} \gamma_{MN} \left(R-\Lambda\right)$.
This formula in rather complicated notations can be found
in \cite{BDW}.

Action (\ref{actiong}) is invariant under general  coordinate transformations
$y^M = y^M(x)$, the corresponding transformation of the metric being
\begin{equation}\label{metrict}
 g^\prime_{RS}(y) \frac{\partial y^R}{\partial x^M}
 \frac{\partial y^S}{\partial x^N} = g_{MN}(x).
\end{equation}
Let us consider infinitesimal coordinate transformations
\beq
\label{trans0}
y^M(x) = x^M + \hat \kappa \xi^M(x).
\eeq
{ Representing the initial and the transformed metrics as}
\bea
g_{MN}(x)& =& \gamma_{MN}(x) + \hat \kappa h_{MN}(x), \nonumber \\
g^\prime_{MN}(y) &=& \gamma_{MN}(y) + \hat \kappa h^\prime_{MN}(y), \nonumber
\end{eqnarray}
substituting these formulas into Eq. (\ref{metrict})
and keeping the terms up to the first order in $ \hat \kappa$, we get the
transformation law for the variation $h_{MN}$,
$$
 h^\prime_{MN}(x) =  h_{MN}(x) - \gamma_{MR}\partial_N  \xi^R -
 \gamma_{NR}\partial_M  \xi^R -  \xi^R \partial_R \gamma_{MN},
$$
which can be equivalently rewritten as
\begin{equation}\label{gaugetr0}
 h^\prime_{MN}(x) =  h_{MN}(x) -\left( \nabla_M \xi_N + \nabla_N \xi_M
 \right).
\end{equation}

It is not difficult to check that the action built with the second
variation Lagrangian (\ref{Lagrangian0}) is invariant under
transformations (\ref{gaugetr0}). Therefore, the latter can be interpreted
as the gauge transformations of the field $h_{MN}$.

Usually the background metric  $\gamma_{MN}$ is a solution of the
Einstein equations
\begin{equation}\label{eq1}
 R_{MN} - \frac{1}{2} \gamma_{MN} \left(R-\Lambda\right) = 8 \pi \hat G
 T_{MN}
\end{equation}
with  a certain energy-momentum tensor of the matter $T_{MN}$.
We express $G_{MN}$ and $\left(R-\Lambda\right)$ in terms of
the energy-momentum tensor  $T_{MN}$ and  substitute it into
(\ref{Lagrangian0}), which gives the second variation Lagrangian
in the form
\begin{eqnarray}\label{Lagrangian01}
L^{(2)}_g/\sqrt{-\gamma} & =& -\frac{1}{4}\left(\nabla_R h_{MN} \nabla^R h^{MN} -
\nabla_R h \nabla^R h + 2\nabla_M h^{MN}\nabla_N h -\right.
 \\ \nonumber
&-& \left. 2 \nabla^R h^{MN} \nabla_M h_{RN}\right) +
\frac{\Lambda}{2(d+2)}\left(h_{MN} h^{MN} -\frac{1}{2} hh\right) -
\\ \nonumber
&-& \frac{4 \pi \hat G}{d+2} T_R^R \left(h_{MN} h^{MN} -\frac{1}{2} hh\right)+
\left(8\pi \hat G T^{MN} h_{MR} h_N^R - 4\pi \hat G T^{MN} h_{MN} h
\right).
\end{eqnarray}
In the next section this expression will be calculated for the
Randall-Sundrum model.

\section{Diagonalization of the  Lagrangian \\in the  Randall-Sundrum model}

The Randall-Sundrum model \cite{RS1} describes the gravity in a
five-dimensional space $E$ with two branes embedded in it (it is often
referred to as the RS1 model). We denote the coordinates as
$\{x^{\mu},x^{4}\}$, $\mu=0,1,2,3$, the coordinate $x^{4}$ parameterizing
the fifth dimension. It forms the orbifold $S^{1}/Z_{2}$, which is the
circle of the circumference $2R$ with points $(x^\mu, x^{4})$ and $(x^\mu,
-x^{4})$ identified. Correspondingly, the metric $g_{MN}$ satisfies the
orbifold symmetry conditions
\begin{equation}
\label{orbifoldsym}
 g_{\mu \nu}(x^\rho,- x^4)=  g_{\mu \nu}(x^\rho,  x^4), \quad
  g_{\mu 4}(x^\rho,- x^4)= - g_{\mu 4}(x^\rho,  x^4), \quad
   g_{44}(x^\rho,- x^4)=  g_{44}(x^\rho,  x^4).
\end{equation}
The branes are located at the fixed points of the orbifold,
$x^{4}=0$ and $x^{4}=R$. In addition we have the usual periodicity
condition which identifies points $(x^\mu, x^{4})$ and
$(x^\mu, x^{4} + 2nR)$.

The action of the model is equal to
\begin{equation}\label{actionRS}
 S = S_g + S_1 + S_2,
\end{equation}
where $S_g$ is given by (\ref{actiong}) with $d=1$ and
\begin{eqnarray}\label{actionsRS}
 S_1&=& V_1 \int_E \sqrt{\tilde g} \delta(x^4) d^5 x,  \\
 S_2&=& V_2 \int_E \sqrt{\tilde g}  \delta(x^4-R) d^5 x.
\end{eqnarray}
{ The subscripts 1 and 2 label the branes.}

The Randall-Sundrum solution for the  metric is {given by}
\begin{equation}\label{metricrs}
  \gamma_{MN} dx^M dx^N = e^{2\sigma} \eta_{\mu\nu} {dx^\mu dx^\nu} + dx^4
dx^4,
\end{equation}
where $\eta_{\mu\nu}$ is the Minkowski metric and  { the function
$\sigma (x^{4})$ in the interval $-R \leq x^{4} \leq R$ is
equal to}
$\sigma(x^4) = -k|x^4|$.
The parameter $k$ has the dimension of mass, and $\Lambda, V_{1,2}$ are
{ related to it as follows:}
$$
\Lambda = -12 k^2, \quad V_1 = -V_2= -\frac{3k}{4\pi \hat G}.
$$
We see that { brane 1}
has positive energy density, whereas { brane 2} has negative one.
The function $\sigma$ has the properties
\begin{equation}\label{sigma}
  \partial_4 \sigma = -k\, sign(x^4), \quad \frac{\partial^2 \sigma}{\partial
  {x^4}^2} =-2k(\delta(x^4) - \delta(x^4-R)) \equiv  -2k\tilde \delta .
\end{equation}

Now we construct the second variation Lagrangian for the  Randall-Sundrum model.
To this end we parameterize the metric in accordance with
(\ref{metricpar}), substitute it into action (\ref{actionRS}) and keep the
terms of the zeroth order in $\hat \kappa$. The contribution of the term
$S_g$ is given by (\ref{Lagrangian01}) with
\begin{equation}\label{energy-mom}
 T_{MN} = -\frac{3k}{4\pi \hat G} \sqrt{\frac{\tilde \gamma}{\gamma}}
 \gamma_{\mu\nu} \delta^\mu_M \delta^\nu_N \tilde \delta,
\end{equation}
${\tilde \gamma}$ denoting the determinant of $ \gamma_{\mu\nu}$.
The contribution of the branes is given by
\begin{equation}\label{branescontrib}
 \Delta L^{(2)}_{1,2} = 3k (h_{\mu\nu} h^{\mu\nu} -\frac{1}{2} \tilde h  \tilde h)
 \tilde \delta \sqrt{-\tilde \gamma},
\end{equation}
where  $\tilde h = \gamma^{\mu\nu} h_{\mu\nu}$.

Thus,  the quadratic Lagrangian for the  variation $h_{MN}$ can be
written as follows:
\begin{eqnarray}\label{Lagrangian}
L/\sqrt{-\gamma} & =& -\frac{1}{4}\left(\nabla_R h_{MN} \nabla^R h^{MN} -
\nabla_R h \nabla^R h + 2\nabla_M h^{MN}\nabla_N h -\right.
 \\ \nonumber
&-& \left. 2 \nabla^R h^{MN} \nabla_M h_{RN}\right)
- 2k^2( h_{MN} h^{MN} -\frac{1}{2}
hh) + \\ \nonumber
& + & \left[ 4k( h_{MN} h^{MN} -\frac{1}{2} hh) + 3k h\tilde h -
6k h_{M\nu} h^{M\nu} + 3k( h_{\mu\nu} h^{\mu\nu} -
\frac{1}{2} \tilde h \tilde h)\right]\tilde \delta.
\end{eqnarray}

It turns out to be convenient to transform the term  $\frac{1}{2}
\nabla^R h^{MN} \nabla_M h_{RN}$ to the standard Fierz-Pauli form
\begin{equation}\label{tranformation}
  \frac{1}{2} \nabla^R h^{MN} \nabla_M h_{RN} = \frac{1}{2}
\nabla_M h^{MN} \nabla^R h_{RN} + \frac{1}{2}
 h^{MN} h^{PQ} R_{MPNQ} - \frac{1}{2}  h^{MN} h_{NP} R_{M}^P,
\end{equation}
where $ R_{MNPQ}$ and $ R_{M}^P$ are the curvature and the Ricci tensors of
the metric $\gamma_{MN}$. If we calculate these terms explicitly, the Lagrangian
becomes
\begin{eqnarray}\label{Lagrangian1}
L/\sqrt{-\gamma} & =& -\frac{1}{4}\left(\nabla_R h_{MN} \nabla^R h^{MN} -
\nabla_R h \nabla^R h + 2\nabla_M h^{MN}\nabla_N h -\right.
 \\ \nonumber
&-& \left. 2 \nabla_M h^{MN} \nabla^R h_{RN}\right)
+\frac{k^2}{2}( h_{MN} h^{MN}+ hh) + \\ \nonumber
& + & \left[ -2k h_{MN} h^{MN} + k h\tilde h -
k h_{M\nu} h^{M\nu} + 3k( h_{\mu\nu} h^{\mu\nu} -
\frac{1}{2} \tilde h \tilde h)\right]\tilde \delta.
\end{eqnarray}

Now let us discuss the gauge invariance of this Lagrangian. The gauge
transformations (\ref{gaugetr0}) in the case under consideration can be
found explicitly and turn out to be
\begin{eqnarray}\label{gaugetrRS}
h'_{\mu\nu}\left(x\right)&=&h_{\mu\nu}\left(x\right)-\left(
\partial_{\mu}\xi_{\nu} +\partial_{\nu}\xi_{\mu}+2\gamma_{\mu\nu}\partial_{4}\sigma\xi_{4}
\right)\\ \nonumber
h'_{\mu4}\left(x\right)&=&h_{\mu4}\left(x\right)-\left(
\partial_{\mu}\xi_{4} +\partial_{4}\xi_{\mu}-2\partial_{4}\sigma\xi_{\mu}
\right)\\ \nonumber
h'_{44}\left(x\right)&=&h_{44}\left(x\right)-2\partial_{4}\xi_{4},
\end{eqnarray}
where the functions $\xi^M(x)$ satisfy the orbifold symmetry conditions
\begin{eqnarray}\label{orbifoldsym1}
\xi^{\mu}\left(x^{\nu},-x^{4}\right)&=&\xi^{\mu}\left(x^{\nu},x^{4}\right)\\
\nonumber
\xi^{4}\left(x^{\nu},-x^{4}\right)&=&-\xi^{4}\left(x^{\nu},x^{4}\right).
\nonumber
\end{eqnarray}
These gauge transformations in other parameterizations were discussed in
papers \cite{CGR} and \cite{AIMVV}.
We will use them to remove the gauge degrees of
freedom of the field $h_{MN}$. To this end we first make a gauge
transformation with
\begin{equation}\label{gauge44}
\xi_{4}\left(x^{\nu},x^{4}\right)=\frac{1}{4}\int_{-x^{4}}^{x^{4}}
h_{44}\left(x^{\nu},y \right)dy -
\frac{x^{4}}{4 R}\int_{-R}^{R}h_{44}\left(x^{\nu},y \right)dy .
\end{equation}
One can  easily see that $\xi_{4}$ satisfies the orbifold symmetry
condition. After this transformation $h_{44}$ takes the form
\begin{equation}
h'_{44}\left(x^{\nu}\right)=\frac{1}{2R}\int_{-R}^{R}
h_{44}\left(x^{\nu},y \right)dy
\end{equation}
and therefore it does not depend on $x^{4}$. Moreover, there are
no remaining gauge transformations with $\xi_{4}$.

Now let us consider  the components $h_{\mu 4}$. Due to orbifold symmetry
\begin{equation}\
h_{\mu4}\left(x,-x^{4}\right)=-h_{\mu4}\left(x,x^{4}\right),
\end{equation}
and the gauge transformations for $h_{\mu4}$ read
\begin{equation}\label{gaugetrMu4}
h'_{\mu4}\left(x,x^{4}\right)=h_{\mu4}\left(x,x^{4}\right)-\left(
\partial_{4}\xi_{\mu}-2\partial_{4}\sigma\xi_{\mu}\right).
\end{equation}
Let us  show that we can impose the condition $h'_{\mu4}=0$. One can
easily find that the corresponding  equation for $\xi_{\mu}$ is
$$
\partial_{4}\left(e^{-2\sigma}\xi_{\mu}\right)=e^{-2\sigma}h_{\mu4}.
$$
Thus, we have
\begin{equation}\label{gaugetrMuNu}
\xi_{\mu}\left(x^{\nu},x^{4}\right)=e^{2\sigma}\int_{0}^{x^{4}}e^{-2\sigma}
h_{\mu4}\left(x^{\nu},y\right)dy,
\end{equation}
and $\xi_{\mu}$ satisfies the orbifold symmetry condition (\ref{orbifoldsym1}).
We will call the gauge with
\begin{equation}\label{unitgauge}
h_{\mu4} =0, \, h_{44} = h_{44}(x^\mu) \equiv \phi (x^\mu)
\end{equation}
the unitary gauge. After  we have  imposed this gauge, there remain gauge
transformations satisfying
\begin{equation}\label{remgaugetr}
\partial_{4}\left(e^{-2\sigma}\xi_{\mu}\right)=0.
\end{equation}
They will be important for removing the gauge degrees of freedom of
the massless mode of the gravitational field.

Varying the action corresponding to the Lagrangian (\ref{Lagrangian1})
we get the following  equations of motion:
\begin{eqnarray}\label{equations}
&\frac{1}{2} \nabla_R  \nabla^R h_{MN} - \frac{1}{2}\gamma_{MN}
\nabla_R  \nabla^R h + \frac{1}{2} \nabla_M  \nabla_N h +
\frac{1}{2}\gamma_{MN} \nabla^R  \nabla^S h_{RS} - &\\ \nonumber
&-\frac{1}{2} \nabla_M  \nabla^R h_{RN} - \frac{1}{2} \nabla_N  \nabla^R
h_{RM}+ k^2(h_{MN} + \gamma_{MN} h) +  \left[-4k h_{MN} - \right. &\\ \nonumber
&\left.  -
k(h_{M\nu}\delta^\nu_N + h_{\mu N}\delta^\mu_M ) + k \gamma_{MN}\tilde h +
k \delta^\mu_M \delta^\nu_N \gamma_{\mu\nu} h +
6k\delta^\mu_M \delta^\nu_N(h_{\mu\nu}-  \frac{1}{2}\gamma_{\mu\nu}\tilde
h)\right]\tilde \delta = 0.&
\end{eqnarray}
Contracting these equations, we get a very simple equation
\begin{equation}\label{contraction}
\frac{3}{2} \nabla^R  \nabla^S h_{RS} - \frac{3}{2} \nabla_R  \nabla^R h +
6 k^2 h - 3k\tilde h \tilde \delta = 0,
\end{equation}
which turns out to be very useful. The equations of motions for different
components are derived from (\ref{equations}) and in the  unitary gauge
(\ref{unitgauge}) take the form:

 1) $\mu\nu$-component
\begin{eqnarray}\label{mu-nu}
 & &\frac{1}{2}\left(\partial_\rho \partial^\rho h_{\mu\nu}-
\partial_\mu \partial^\rho
h_{\rho\nu}-\partial_\nu \partial^\rho h_{\rho\mu} + \frac{\partial^2
h_{\mu\nu}}{\partial {x^4}^2}\right)- 2k^2  h_{\mu\nu} +
\frac{1}{2}\partial_\mu \partial_\nu \tilde h +
\frac{1}{2}\partial_\mu \partial_\nu \phi+  \\ \nonumber
&+ & \frac{1}{2} \gamma_{\mu\nu}\left(\partial^\rho \partial^\sigma h_{\rho\sigma} -
\partial_\rho \partial^\rho \tilde h -  \frac{\partial^2
\tilde h}{\partial {x^4}^2}-4\partial_4 \sigma \partial_4 \tilde h
 - \partial_\rho \partial^\rho \phi + 12 k^2 \phi\right)  \\ \nonumber
&+& \left[2k  h_{\mu\nu} - 3k\gamma_{\mu\nu}\phi \right]\tilde
\delta = 0
\end{eqnarray}

 2) $\mu 4$-component,
\begin{equation}\label{mu-4}
\partial_4 ( \partial_\mu \tilde h - \partial^\nu  h_{\mu\nu})-
3\partial_4 \sigma \partial_\mu \phi = 0,
\end{equation}
which plays the role of a constraint,

 3) $4 4$-component
\begin{equation}\label{4-4}
\frac{1}{2}(\partial^\mu \partial^\nu  h_{\mu\nu} - \partial_\mu
\partial^\mu \tilde h ) - \frac{3}{2}\partial_4 \sigma \partial_4 \tilde h
+ 6 k^2 \phi =0.
\end{equation}

The contracted equation can also be rewritten in this gauge to be
\begin{equation}\label{contracted}
\frac{3}{2}(\partial^\mu \partial^\nu  h_{\mu\nu} - \partial_\mu
\partial^\mu \tilde h ) - \frac{15}{2}\partial_4 \sigma \partial_4 \tilde h
-  \frac{3}{2}\frac{\partial^2 \tilde h}{\partial {x^4}^2} -
\frac{3}{2}\partial_\mu \partial^\mu \phi  + 30 k^2 \phi -
12 k \phi \tilde \delta =0.
\end{equation}

Let us first consider the $44$-component and the contracted  equation.
Multiplying { Eq. (\ref{4-4})} by 3 and subtracting it from
{ Eq. (\ref{contracted})},
we { obtain the following} equation, containing $\tilde h$ and $\phi$ only:
\begin{equation}\label{contracted-44}
\frac{\partial^2 \tilde h}{\partial {x^4}^2}  + 2\partial_4 \sigma \partial_4 \tilde h
  -8k^2 \phi+ 8k \phi \tilde \delta + \partial_\mu \partial^\mu \phi = 0.
\end{equation}

To describe correctly the physical degrees of freedom in the
model we write the multidimensional gravitational field as
\beq\label{substitution}
 h_{\mu\nu} =  b_{\mu\nu} + \gamma_{\mu\nu}(\sigma - c)\phi +
 \frac{1}{2k^2} \left(\sigma - c +\frac{1}{2} +
 \frac{c}{2}e^{-2\sigma}\right) \partial_\mu \partial_\nu \phi   ,
\eeq with $\sigma=\sigma(x^4)$ and $c$ being a constant. We will see that
the field $b_{\mu\nu}(x^{\mu},x^{4})$ describes a massless graviton
\cite{RS1,RS2} and massive Kaluza-Klein spin-2 fields, whereas $\phi
(x^{\mu})$ describes a scalar field called radion. Apparently, the radion
field was first identified in Ref. \cite{CGR}. As it will be shown,
relation (\ref{substitution}) with an appropriately chosen constant $c$
diagonalizes the second variation Lagrangian  and decouples the equations
of motion (\ref{mu-nu})-(\ref{4-4}). The form of the substitution
(\ref{substitution}) is suggested by the form of the gauge transformation
that transforms the theory from the local Gaussian normal coordinates (i.e.
coordinates corresponding to $g_{44}=1$, $g_{\mu4}=0$) \cite{CGR}, where
the second variation Lagrangian is diagonal in the bulk by construction, to
our coordinates. We would like to note that in fact the equations of motion
derived in Ref. \cite{CGR} remain coupled at the fixed points of the
orbifold, i.e. at the points where the branes are located. To decouple them
the term $\sim e^{-2\sigma}$ in Eq. (\ref{substitution}) is needed. It
follows from this expression that
\beq \label{substitution1} \tilde h =
 \tilde b + 4 (\sigma -c)\phi+
 \frac{1}{2k^2} \left(\sigma - c +\frac{1}{2} +
 \frac{c}{2}e^{-2\sigma}\right) \partial_\mu \partial^\mu \phi.
\eeq

Substituting  (\ref{substitution1}) into Eq. (\ref{contracted-44}) we get
the equation
 $$
\partial_4(e^{2\sigma}\partial_4 \tilde b) +
2 c \partial_\mu \partial^\mu \phi +\frac{2}{k}(\sigma -c +c
\,e^{-2\sigma}) \partial_\mu \partial^\mu \phi \, \tilde \delta =0.
 $$
 Choosing the constant $c$ so that the boundary term, proportional to
$\tilde{\delta} (x^{4})$, vanishes, i.e.
\begin{equation}\label{constant}
  c= \frac{kR}{e^{2kR} -1},
\end{equation}
we reduce the equation to the following one:
\begin{equation}\label{difference1}
\partial_4(e^{2\sigma}\partial_4 \tilde b) +
2 c \partial_\mu \partial^\mu \phi  =0.
\end{equation}
Next, taking into account that $\gamma_{\mu\nu} =
e^{2\sigma}\eta_{\mu\nu}$, we pass to the flat metric in this equation
 and denote $ b = \eta^{\mu\nu} b_{\mu\nu}$.
Then we get
\begin{equation}\label{relation3}
\partial_4 ( e^{2\sigma}\partial_4 (e^{-2\sigma}b))+
2 c e^{-2\sigma} \partial_\mu \partial^\mu \phi  =0.
\end{equation}

Let us consider Fourier expansions of both terms in $x^4$. Since the term
with the derivative $\partial_4$ has no zero modes, this equation implies
that the radion field is a free massless field \cite{CGR}, i.e
\begin{equation}\label{radioneq}
 \partial_\mu \partial^\mu \phi = 0,
\end{equation}
and $\partial_4 (e^{-2\sigma}b) = e^{-2\sigma} f(x^\mu)$. Applying the same
reasoning to the last relation, we get
$$
\partial_4 (e^{-2\sigma}b) = 0.
$$

Recall that we have at our disposal the gauge transformations,
satisfying  (\ref{remgaugetr}). With the help of these
transformations, we can impose  the gauge
\beq \label{T}
\tilde b = b =0,
\eeq
{ It is easy to see that} there remain gauge transformations
{  parameterized} by
$\xi_\mu = e^{2\sigma}\epsilon_\mu(x^\nu)$ with $\epsilon_\mu(x^\nu)$
satisfying $\quad \partial^\mu \epsilon_\mu = 0$.
Substituting expression
(\ref{substitution}) into Eqs. (\ref{mu-4}), (\ref{4-4}) and
passing to gauge (\ref{T}), we arrive at the following system
{ of relations}:
\begin{eqnarray}
\label{system3}
\label{equ1}
\partial^\mu \partial^\nu  b_{\mu\nu}& =&0, \\
\label{equ2}
\partial_4 (e^{-2\sigma} \partial^\mu  b_{\mu\nu})&=&0.
\end{eqnarray}
The remaining gauge transformations are sufficient to { impose the
condition}
\begin{equation}\label{gaugecond}
 \partial^\mu  b_{\mu\nu} = 0.
\end{equation}
{ The conditions (\ref{T}) and (\ref{gaugecond}) define
the gauge often called the transverse-traceless (TT) gauge.
Having imposed this gauge, we are still left with residual}
gauge transformations
\begin{equation}
\label{gaugetr}
\xi_\mu = e^{2\sigma}\epsilon_\mu(x^\nu), \quad
\partial_\rho \partial^\rho \epsilon_\mu = 0,
\quad \partial^\mu \epsilon_\mu = 0,
\end{equation}
which, { as we will see shortly}, are important for
determining the number of degrees of freedom of the
massless mode of $b_{\mu\nu}$.

{ In the gauge (\ref{T}), (\ref{gaugecond})}
the constraint (\ref{mu-4}) and the $44$-equation
(\ref{4-4}) { are trivially satisfied. Let us turn
to Eq. (\ref{mu-nu}).}
{ Substituting expressions (\ref{substitution}), (\ref{substitution1})
into it and passing to TT gauge,
we transform this equation to the following form}
\begin{equation}\label{mu-nu1}
  \frac{1}{2} \partial_\rho \partial^\rho b_{\mu\nu} +
  \frac{1}{2}  \frac{\partial^2 b_{\mu\nu}}{\partial {x^4}^2} -
  2k^2 b_{\mu\nu} +  2k b_{\mu\nu} \tilde \delta= 0.
\end{equation}
Thus, we have decoupled all the equations.

{ To summarize,
fluctuations around the Randall-Sundrum solution are described by the
spin-2 field $b_{\mu \nu}(x^{\mu},x^{4})$ and the massless radion field
$\phi (x^{\mu})$. Their classical equations of motion are given by Eq.
(\ref{mu-nu1}) and Eq. (\ref{radioneq}) respectively.}

{ Now let us consider the second variation action of the theory.
Substituting} (\ref{substitution}), (\ref{substitution1}) with  $c$ given by
(\ref{constant}) into the Lagrangian (\ref{Lagrangian1}) and  taking into
account the TT gauge conditions { (\ref{T}), (\ref{gaugecond})}
for $b_{\mu\nu}$, we { obtain}
\begin{equation}\label{Lagrangian3}
  L/\sqrt{-\gamma} = -\frac{1}{4} \partial_\rho  b_{\mu\nu} \partial^\rho
  b^{\mu\nu}  -\frac{1}{4}( \partial_4  b_{\mu\nu} -2\partial_4 \sigma
  b_{\mu\nu})( \partial_4  b^{\mu\nu} +2\partial_4 \sigma
  b^{\mu\nu}) - \frac{3}{4}  c\,  e^{-2 \sigma}\partial_\mu \phi\partial^\mu
  \phi.
\end{equation}

 To get an effective Lagrangian for the system, it remains to decompose
the field $b_{\mu\nu}$ into the modes with
 definite masses and to integrate $L$, Eq. (\ref{Lagrangian3}), over $x^4$.
 After integration over $x^4$ we obtain the
{ following four-dimensional} effective Lagrangian for the field $\phi$:
\begin{equation}\label{Lagrangian_phi}
L_\phi = -\frac{3}{2}\frac{kR^2}{ e^{2kR}-1}  \partial_\mu \phi\partial^\mu
\phi,
\end{equation}
where the index is raised with the flat metric.
 To bring the kinetic term to the canonical form we  have to rescale the
field according to
\begin{equation}\label{rescaling_phi}
 \phi \rightarrow \sqrt{\frac{ e^{2kR}-1}{3 kR^2}} \phi.
\end{equation}

{ Let us turn to the mode expansion of the field $b_{\mu \nu}$.}
In what follows we denote $x^4 =y$. Following Refs. \cite{RS2,DHR1,DHR2},
to  perform the { expansion we first find eigenfunctions $\psi_{n}(y)$
and eigenvalues $m_{n}$ of the problem}
\begin{equation}
\label{bessel0}
 \left[\frac{1}{2} \frac{d^2}{dy^2} +
 2k(\delta(y) - \delta(y-R)) - 2k^2 \right] \psi_{n}(y) =
 \frac{m_{n}^{2}}{2} e^{2k|y|} \psi_{n}(y).
\end{equation}
{The operator in the l.h.s. is a part of the equation
(\ref{mu-nu1}). Eq. (\ref{bessel0})}
can be solved exactly. To this end we first introduce the  variable
$$
\xi = \frac{m_{n}}{k} e^{k|y|}.
$$
{ Note that the eigenvalue $m_{n}$ enters into the definition of
$\xi$.}
{ Making the change of variables} in Eq.
(\ref{bessel0}) { we transform it to}
\begin{equation}\label{{bessel1}}
\left[ \frac{d^2}{d\xi^2} + \frac{1}{\xi} \frac{d}{d\xi} + 1 -
\frac{4}{\xi^2} + \frac{4}{\xi}\left(\delta(\xi - \xi_1) - \delta(\xi -
\xi_2)\right)\right]f(\xi) = 0,
\end{equation}
 where $\xi_1 = m_n/k$ and $\xi_2 =  m_n/k \exp(kR)$. This equation without
 the $\delta$-function
 terms is just the Bessel equation, the general solution being a linear
 combination of Bessel and Neumann functions $J_2(\xi)$ and $Y_2(\xi)$:
 $$
Z_2(\xi) = a J_2(\xi) + b Y_2(\xi).
 $$
The term with $\delta$-functions can be taken into account by imposing the
boundary condition
$$
Z_2^\prime(\xi) + \frac{2}{\xi} Z_2(\xi) = 0
$$
at $\xi =\xi_1 = m_n/k$ and $\xi = \xi_2 =  m_n/k \exp(kR) $.
The first boundary condition, at $\xi = \xi_{1}$, can be satisfied by an
appropriate choice
of  the coefficients $a$ and $b$:
\begin{equation}\label{besselZ}
Z_2(\xi) = Y_1 \left(\frac{m_n}{k}\right) J_2(\xi) -
J_1 \left(\frac{m_n}{k}\right) Y_2(\xi).
\end{equation}
 The second boundary condition, at $\xi = \xi_{2}$, defines
the mass spectrum of the theory and can be rewritten as
\begin{equation}\label{eigenvalues1}
    J_1(\xi_1)Y_1(t\xi_1) - J_1(t\xi_1) Y_1(\xi_1) =0,
\end{equation}
where $ t= \exp(kR)$. There exists a theorem
about such combinations of products of Bessel and Neumann functions,
which asserts that for  $t > 1$ this combination is
an even function of $x$ and its zeros are real and simple \cite{BE}.
Thus, the non-zero masses of tensor gravitons are defined by the positive
zeros $\gamma_n, \, n=1,2, \cdots \,\, $ of this combination  and explicitly given
by the  formula
\begin{equation}\label{zerosb}
  m_n = \gamma_n k \equiv \beta_n  k exp(-kR).
\end{equation}
As it was discussed in many papers (see Refs. \cite{RS1,RS2,DHR1,DHR2}),
{ the product $kR$ is chosen to
be of the order $30 \div 35$ in order to solve the hierarchy problem.}
{Since $exp(-kR)$ is a tiny factor, the ratio $m_{n}/k \ll 1$
for eigenvalues with small $n$, such that $\beta_{n} \sim {\cal O}(1)$.
Using the expansions for
the Bessel functions for the small argument it is easy to see that
such $\beta_{n}$ are approximately given by the roots of the equation
$J_{1}(\beta_{n}) = 0$.}

Let {us describe the complete orthonormal system
$\{\psi_n(y), n =0,1, \cdots \}$}
of solutions of equation (\ref{bessel0}), satisfying
$$
\int_{-R}^R e^{2k|y|} \psi_m (y) \psi_n (y)dy = \delta_{mn}.
$$
The normalized functions  $\psi_n(y)$ corresponding to the mass
eigenvalue  $m_0 = 0$ and non-zero masses $m_n = \gamma_{n} k > 0$ are given by
\begin{eqnarray}\label{eigenfunctions}
\label{zeromass}
 \psi_0 (y) & = & N_{0}\, e^{-2k|y|}, \quad
 N_{0}= \frac{ k^{\frac{1}{2}}}{(1-\exp{(-2kR)})^{\frac{1}{2}}} \\
 \nonumber
 \psi_n (y)& = &N_n \left( Y_1 \left(\gamma_{n}\right)
  J_2 \left(\gamma_{n} e^{k|y|}\right) -
J_1 \left(\gamma_{n}\right) Y_2 \left( \gamma_{n}
e^{k|y|}\right)\right)\\
 N_{n}&= &\frac{ k^{\frac{1}{2}}}{(\exp(2kR)Z_2^2(\gamma_{n}\exp(kR))-
 Z_2^2(\gamma_{n}) )^{\frac{1}{2}}}, \quad \gamma_{n} = \frac{m_n}{k} = \beta_n e^{-kR},
\end{eqnarray}
where $Z_2$ was defined in (\ref{besselZ}). Now we calculate the
values of the eigenfunctions for small $n$ at $y=0$ and $y=R$,
which will be used later. Taking into account that $\exp (kR) \gg 1$
we obtain the following expressions:
\bea
  \psi_{0} (0) & = & N_{0} \approx \sqrt{k}; \; \; \;
  \psi_{n} (0) \approx \frac{\sqrt{k}}{|J_{2}(\beta_{n})|} e^{-kR}, \label{psi-0} \\
 \psi_{0} (R) & = & N_{0} e^{-2kR} \approx \sqrt{k} e^{-2kR}; \; \; \;
  \psi_{n} (R) \approx - \sqrt{k} e^{-kR}. \label{psi-1}
\eea

 Finally, we expand { $b_{\mu\nu}(x,y)$ as follows}:
\begin{equation}\label{decomp}
b_{\mu\nu}(x,y) = \sum_n b^n_{\mu\nu}(x) \psi_n(y),
\end{equation}
$ b^n_{\mu\nu}(x)$ being ordinary spin-2 transverse traceless
4-dimensional  fields, which can be
quantized by the standard procedure. Substituting this decomposition into
(\ref{Lagrangian3}), passing to the corresponding action, integrating explicitly
over the orbifold  and rescaling the radion field $\phi$ in accordance with
(\ref{rescaling_phi}), we get an effective action for the physical degrees
of freedom of the 5-dimensional gravitational field
\begin{equation}\label{effaction}
S_{eff} = \int\left( -\frac{1}{4}\sum_n(\partial_{\mu}b^n_{\rho\sigma}
\partial^{\mu}b^{n,\rho\sigma}+
{m_n^{2}} b^n_{\rho\sigma}b^{n,\rho\sigma})
-\frac{1}{2} \partial_{\mu}\phi \partial^{\mu}\phi\right) dx.
\end{equation}

 Thus,  we have a massless spin-2 tensor field, an infinite tower of massive
spin-2 tensor  fields and  just one massless  scalar radion. A very important
point is that, due to  the form of the solution with zero mass (\ref{zeromass}),
the  remaining gauge  transformations  (\ref{gaugetr}) result in the following
gauge transformations  of  the field  $ b^0_{\mu\nu}(x)$:
\begin{equation}\label{gaugetr0m}
  b^{\prime 0}_{\mu\nu}(x)=  b^0_{\mu\nu}(x) -(\partial_\mu \zeta_\nu(x)+\partial_\nu
  \zeta_\mu(x)), \quad \partial^\mu \zeta_\mu(x) =0, \quad \partial_\mu \partial^\mu
  \zeta_\nu(x) =0.
\end{equation}
This   guarantees that the field $ b^0_{\mu\nu}(x)$ has  only two degrees of freedom,
and, therefore, can be identified with the field of the massless graviton.

\section{Interaction with matter on the branes}

  Now we have to find the interaction of these fields
 with the matter on the branes. The general form of this interaction is
 standard,
\begin{equation}\label{interaction}
  \frac{\hat\kappa}{2} \int_{B_1} h^{\mu\nu}(x,0) T^1_{\mu\nu} dx +
   \frac{\hat\kappa}{2} \int_{B_2} h^{\mu\nu}(x,R) T^2_{\mu\nu} \sqrt{- \tilde{\gamma}(R)} dx,
\end{equation}
where $ T^1_{\mu\nu}$ and $ T^2_{\mu\nu}$ are energy-momentum tensors of the
matter on brane 1 and  brane 2 respectively:
$$
 T^{1,2}_{\mu\nu} =2 \frac{\delta L^{1,2}}{\delta  \gamma^{\mu\nu}} -
  \gamma^{1,2}_{\mu\nu}  L^{1,2}.
$$
Substituting (\ref{substitution}) into (\ref{interaction}), decomposing
$b_{\mu\nu}(x,y)$ according to (\ref{decomp}) and rescaling the field $\phi$
according to (\ref{rescaling_phi}), we find that the interaction of the
graviton fields $b^n_{\mu\nu}(x)$ and the massless canonically normalized
radion field $\phi$ with the matter on  brane 1 is given by
\begin{equation}\label{intb}
 \frac{1}{2}\int_{B_1} \left( \kappa_1   \, b^0_{\mu\nu}(x)
 T^{\mu\nu}+ \kappa_1 \sum_{n =1}^\infty  \frac{\psi_n (0)}{N_{0}}\, b^n_{\mu\nu}(x)
 T^{\mu\nu}-\frac{\kappa_2}{ \sqrt{3}}  \phi\, T_\mu^\mu\right)dx.
\end{equation}

Here
\beq
\kappa_1 = \hat \kappa N_0, \; \; \;
\kappa_2 = \kappa_1 e^{-kR}  \label{kappa-def}
\eeq
(the normalization factor $N_{0}$ was
defined in (\ref{eigenfunctions})).
Since  $\kappa_1$  is the coupling constant of the massless graviton,
it can be  expressed in terms of the  4-dimensional gravitational
constant $G_1$ on   brane 1 as $\kappa_1 = \sqrt{16 \pi G_1}$,
which gives the relation
$$
  G_1 = \frac{\hat G k}{1-e^{-2kR}}
$$
between the constants on brane 1 \cite{RS1,Rub01}.

To find the interaction of the 5-dimensional gravity with the matter on
  brane 2 turns out to be a more complicated problem, because the
coordinates $\{x^\mu\}$ in (\ref{metricrs}), in  which we work,  are not
Galilean on this brane (coordinates are called Galilean, if
$g_{00}= -1,\, g_{11}=g_{22}=g_{33} =1$, see \cite{LL}).

To solve this problem, we introduce the  Galilean coordinates $\{z^\mu\}$
on  brane 2 related to $\{x^\mu\}$ by $x^\mu = e^{kR}z^\mu$. When we
do this in (\ref{interaction}), we get for   brane 2
\begin{equation}\label{intf1}
 \frac{\hat \kappa}{2} \int_{B_2} h^{\prime}_{ \mu\nu}(z,R) T^{\prime 2,\mu\nu} dz,
\end{equation}
where $T^{\prime 2,\mu\nu}$ is the canonical  energy-momentum tensor of
the matter, $ h^{\prime}_{ \mu\nu}(z,R)= e^{2kR} h_{\mu\nu}(x,R)$, and
the  indices  are raised with the flat metric. Thus, we see
that in order to find the interaction of  the 5-dimensional gravity with
the matter on   brane 2 we have to pass  to the Galilean
coordinates  $\{z^\mu\}$ in the effective action (\ref{effaction}) as well.
This results in the following expression:
\begin{equation}\label{effaction1}
S_{eff} = \int\left( -\frac{1}{4}\sum_n( e^{-2kR}\partial_{\mu}
b^{\prime n}_{\rho\sigma}\partial^{\mu}b^{\prime n,\rho\sigma}+
{m_n^{2}} b^{\prime n}_{\rho\sigma}b^{\prime n,\rho\sigma})
-\frac{1}{2}  e^{2kR}\partial_{\mu}\phi^\prime
\partial^{\mu}\phi^\prime\right) dz,
\end{equation}
where the indices  are also raised with the flat metric. To
transform this expression to be a canonical action we have to rescale the
fields $b^{\prime n}_{\rho\sigma},\,\phi^\prime $ and the masses $m_n$.
We define the  graviton fields $u^n_{\rho\sigma} = e^{-kR}
b^{\prime n}_{\rho\sigma}$
and the radion field $\varphi= e^{kR}\phi^\prime$, as  seen from
 brane 2, for which the expression (\ref{effaction1}) reduces
to a canonical action in  the Galilean coordinates  $\{z^\mu\}$:
\begin{equation}\label{effaction2}
S_{eff} = \int \left( -\frac{1}{4}\sum_n(\partial_{\mu}u^n_{\rho\sigma}
\partial^{\mu}u^{n,\rho\sigma}+
{(m_n e^{kR})^{2}} u^n_{\rho\sigma}u^{n,\rho\sigma})
-\frac{1}{2} \partial_{\mu}\varphi \partial^{\mu}\varphi\right) dz.
\end{equation}
Now we are able to write down the interaction of these fields
with the matter on  brane 2, which is found to be
\begin{equation}\label{intf2}
 \frac{1}{2}\int_{B_2} \left( \kappa_2 \,   u^0_{\mu\nu}(x)
 T^{\mu\nu}+\kappa_2 \sum_{n=1}^\infty  \frac{\psi_n (R)e^{2kR}}{N_{0}}\,
  u^n_{\mu\nu}(x)
 T^{\mu\nu}-\frac{\kappa_1}{ \sqrt{3}}  \varphi\, T_\mu^\mu\right)dz.
\end{equation}
The coupling constants  $\kappa_1$ and $\kappa_2$ are  defined  by
Eqs. (\ref{kappa-def}). Since   $\kappa_2$ is the coupling  constant
of the massless graviton, formally it
 gives the  following relations between the 5-dimensional
gravitational constant $\hat G$ and the four-dimensional gravitational
constant $G_2$ on  brane 2
\beq\label{G2-G}
  G_2 = \frac{\hat G k}{e^{2kR}-1},
\eeq
which coincides with the one  found in \cite{Rub01}.

\section{Conclusions and discussion}

In the present paper we have developed a Lagrangian description of
linearized gravity in Randall-Sundrum model with two branes, which
enabled us to find easily the physical degrees of
freedom of this model and to construct an effective Lagrangian for them.
If we take the limit $R\rightarrow \infty$, the radion field $\phi$ drops
from the Lagrangian, and we get the same degrees of freedom, as found
in \cite{AIMVV} for the case of infinite extra dimension.

A very important point of our study is the observation that the
5-dimensional gravity looks different, when viewed from different
branes: the masses of the gravitational KK modes and the coupling
constants   of the massless fields differ in the exponential
factor.

The  case of finite  extra dimension is more interesting from the
point of view  of  possible phenomenology  scenarios due to  the
presence of the radion field and of two coupling constants,
 $\kappa_1$ and $\kappa_2$. These scenarios  will be discussed
 in a separate paper.  Now we only mention two possibilities.

Let us assume that our brane is the  brane with the positive energy
density, in contrast to the standard  RS1  scenario, so that
the coupling of the massless mode $\kappa_{1} \sim 1/M_{Pl}$. If we make a
natural assumption  that $kR \gg 1$, then $\kappa_1 \gg \kappa_2$, and we see
from (\ref{psi-0}), (\ref{intb}) that the interactions of the massive KK
modes and of the radion field with the matter on our  brane are much weaker
(exponentially suppressed) than that of the massless graviton. Therefore,
the radion field does not  affect  Newton's gravity on our brane, which
arises from the  interaction of the massless graviton in (\ref{intb}). The
massive KK modes have masses of the order of $M_{pl} e^{-kR}$.

If one assumes that our brane is  brane 2
with the negative energy density, then, as follows from (\ref{intf2}),
the interaction of massless radion field is not suppressed.
In this case the constant $G_{2}$ can be identified with
Newton's constant, $G_{2} = 1/M_{Pl}^{2}$. The relation (\ref{G2-G}) gives the
following relation between the Planck mass and the fundamental
mass scale $M$ in the theory defined as $\hat{G}=1/M^{3}$:
$$
M_{Pl}^{2} = \frac{M^{3}}{k} \left( e^{2kR}-1 \right) \approx
\frac{M^{3}}{k} e^{2kR}.
$$
{ (see \cite{Rub01}).
By choosing, for example, $M \sim k \sim 1$TeV we reproduce the
correct value of the Planck mass, if the argument of the exponential factor
satisfies $kR \approx 30 \div 35$. Then, as in the standard scenario,
the couplings of the massive KK states and of the radion are of the order
of $1 \; \mbox{TeV}^{-1}$ \cite{DHR1,DHR2}. This can be
easily seen from Eqs. (\ref{psi-1}), (\ref{intf2}). The masses of the
KK excitations, which can be read from Eqs.  (\ref{zerosb}), (\ref{effaction2}),
are of the order of $1 \; \mbox{TeV}$ for small $n$.
The presence of the massless radion with such coupling
leads to some predictions which
are in contradiction with the available high energy physics data.
To solve this problem, several mechanisms for generating a mass for the
radion  were proposed (see, for example, Ref. \cite{wise}).

\bigskip
{ \large \bf Acknowledgments}
\medskip

The authors are grateful to G.Yu. Bogoslovsky, E.R. Rahmetov and V.A.
Rubakov  for useful discussions.  The work of E.B., Yu.K. and I.V. was
supported by RFBR grant 00-01-00704  and the grant 990588 of the programme
"Universities of Russia".
 E.B. was supported in part by the CERN-INTAS
grant 99-0377 and by the INTAS grant 01-0679.
 Yu.K. and I.V. were also supported
in part by the programme SCOPES (Scientific co-operation between
Eastern Europe and Switzerland) of the Swiss National Science
Foundation, project No. 7SUPJ062239, and financed by Federal
Department of Foreign Affairs.

\bigskip
{ \large \bf Note added}
\medskip

We decided to replace the text, because many misprints were found
during these years. All of them have been corrected in the new
version. We would like to thank M.~Iofa, who pointed to us some of
the misprints.

\end{document}